\documentclass[]{spie}  %>>> use for US letter paper
\usepackage[]{graphicx}
\usepackage{color} 
\usepackage{amsmath}
\usepackage{verbatim}
\usepackage{hyperref}   % use for hypertext links, including those to external documents and URLs
\usepackage{array}
\usepackage{amssymb}

\title{Optimization of $\leq$ 200$\mu$m pitch CZT detectors for future high-resolution 
X-ray instrumentation in astrophysics} 

\author{Anna Zajczyk\supit{a}, Marie Draper\supit{a}, Paul Dowkontt\supit{a},
Qingzhen Guo\supit{a}, Fabian Kislat\supit{a}, Henric Krawczynski\supit{a}, 
Gianluigi De Geronimo\supit{b}, Shaorui Li\supit{b}, Matthias Beilicke\supit{a}
\skiplinehalf
\supit{a}Department of Physics and McDonnell Center for Space Sciences, Washington University in St. Louis, 1 Brookings Drive,
St. Louis, MO 63105, USA; \\
\supit{b}Brookhaven National Laboratory, Upton, NY 11973, USA;\\
}

\authorinfo{Further author information: (Send correspondence to A. Zajczyk): azajczyk@physics.wustl.edu}
\begin{document} 
  \maketitle 

%%%%%%%%%%%%%%%%%%%%%%%%%%%%%%%%%%%%%%%%%%%%%%%%%%%%%%%%%%%%% 
\begin{abstract}
Cadmium Zinc Telluride and Cadmium Telluride are the detector materials of choice for the
detection of X-rays in the X-ray energy band $E \ge 5\, \rm{keV}$ with excellent
spatial and spectral resolution and without cryogenic cooling. Owing to
recent breakthroughs in grazing incidence mirror technology,
next-generation hard X-ray telescopes will achieve angular resolution
between 5 and 10 arc seconds - about an order of magnitude better
than that of the NuSTAR hard X-ray telescope. As a consequence, the next
generation of X-ray telescopes will require pixelated X-ray
detectors with pixels on a grid with a lattice constant of $\leq 250\, \mu$m.
Additional detector requirements include a low energy threshold of less than $5\, \rm{keV}$
and an energy resolution of less than one keV. The science drivers for a
high angular-resolution X-ray mission include studies and
measurements of black hole spins, the cosmic evolution of
super-massive black holes, active galactic nuclei feedback, and the behaviour of
matter at very high densities. In this contribution, we report on our R\&D studies 
with the goal to optimise small-pixel Cadmium Zinc Telluride and Cadmium Telluride detectors.
\end{abstract}

%>>>> Include a list of keywords after the abstract 


%%%%%%%%%%%%%%%%%%%%%%%%%%%%%%%%%%%%%%%%%%%%%%%%%%%%%%%%%%%%%
\section{INTRODUCTION}
\label{sec:intro}  % \label{} allows reference to this section

Owing to recent breakthroughs in grazing incidence mirror technology,
next-generation X-ray telescopes will achieve angular resolution
of $5-10$ arc seconds \cite{XrayMirrorsNew_Zhang_SPIEE,
XrayMirrorsNew_Ramsey}. Such telescopes require pixelated detectors with a
pixel-to-pixel pitch of $\leq 250\, \mu \rm{m}$. A low energy threshold of $<5
\, \rm{keV}$ and an energy resolution of $<1 \, \rm{keV}$ are
additional requirements for future X-ray satellite missions to enable
spectroscopic studies of emission processes in the universe.
The science drivers for a high angular-resolution X-ray mission
include studies and measurements of black hole spins, the cosmic
evolution of super-massive black holes, active galactic nuclei (AGN) feedback, 
and the behaviour of matter at very high densities.

A high stopping power, good spectral performance and good imaging capabilities
are required from materials used as detectors in the X-ray energy band. 
Semiconductors like silicon (Si) and germanium (Ge) yield excellent 
energy resolution. However, due to its low density of 2.33 g cm$^{-3}$ and low atomic number of 14,
silicon is inefficient at stopping high energy photons, which limits 
its use to only low energies. At the same time, due to its small band gap
germanium needs to be operated at cryogenic temperatures in order to decrease 
its dark current. Thus, semiconductors with high atomic number $Z$, high density 
$\rho$ and wide band gap $E_{\rm gap}$ that meet the aforementioned requirements have
been developed. Refer to Tab.~\ref{tab:semiconductors} for comparison of 
properties of selected semiconductors.

\begin{table}
\begin{center}
	\caption{\small Properties of selected semiconductors\cite{cdte2003}: density $\rho$, 
	atomic number $Z$, band gap width $E_{\rm gap}$, and intrinsic energy
	resolution $\Delta E_{\rm intr}$. The width of the band gap of CZT crystal 
	increases with the increase of Zn concentration\cite{cdte2001}. The widely used composition
	of Cd$_{1-x}$Zn$_{x}$Te has $x =0.08 -0.15$ resulting in the band gap of 
	$\sim1.6\, \rm{eV}$.}
	\begin{tabular}{ccccc}
	\multicolumn{5}{c}{}\\ \hline
	Semi-     & $\rho$ & $Z$ & $E_{\rm gap}$ & $\Delta E_{\rm intr}$ \\ 
	conductor & [g cm$^{-3}$] &  & [eV] & [eV] \\ \hline \hline
	Si & 2.33 & 14 & 1.12 & 450 \\
	Ge & 5.33 & 32 & 0.67 & 400 \\
	CdTe & 5.85 & 48, 52 & 1.44 & 700\\
	CZT & 5.81 & 48, 52 & 1.6-2.2 & 620\\ \hline
	\end{tabular}
\end{center}
\label{tab:semiconductors}
\end{table}

Cadmium telluride (CdTe) and cadmium zinc telluride (CZT) crystals 
proved to be a well-suited material for detecting hard X-ray and soft $\gamma$-ray 
photons in astrophysical applications (e.g. Burst Alert Telescope\cite{bat2004} 
on board the Swift satellite consists of 32,768 CZT detectors sensitive to photons 
in the energy range between $15\, \rm{keV}$ and $150\, \rm{keV}$). Properties that make CdTe/CZT
almost perfect X-ray detector materials are their high average atomic numbers
($Z_{\rm Cd} = 48$, $Z_{\rm Te} = 52$) and their high densities ($\rho_{\rm CdTe} = 
5.85$ g cm$^{-3}$, $\rho_{\rm CdZnTe} = 5.81$ g cm$^{-3}$). The former causes 
photoelectric absorption to be a dominant photon-interaction process in 
the material up to about $300\, \rm{keV}$. The latter results in a good photon stopping
power. Moreover, a wide band gap of CdTe/CZT allows for operation at room
temperature. Though not devoid of problems of their own\cite{cdte2001, cdte2003}, 
e.g. charge trapping or ballistic deficit, CdTe/CZT crystals have an intrinsic
energy resolution\footnote{Intrinsic energy resolution is the energy resolution
derived from taking into account only statistical fluctuations of electron-hole pair 
creation.} comparable to that of germanium. The presented properties of CdTe/CZT
and our prior experience in fabrication of and signal processing from CZT detectors 
make these semiconductors our materials of choice for performance studies of fine-pitch 
$\leq 350\, \mu$m pixelated anodes for future high-resolution X-ray detectors.

\begin{figure}
  \begin{center}
  \begin{tabular}{cc}
    \includegraphics[scale=0.52]{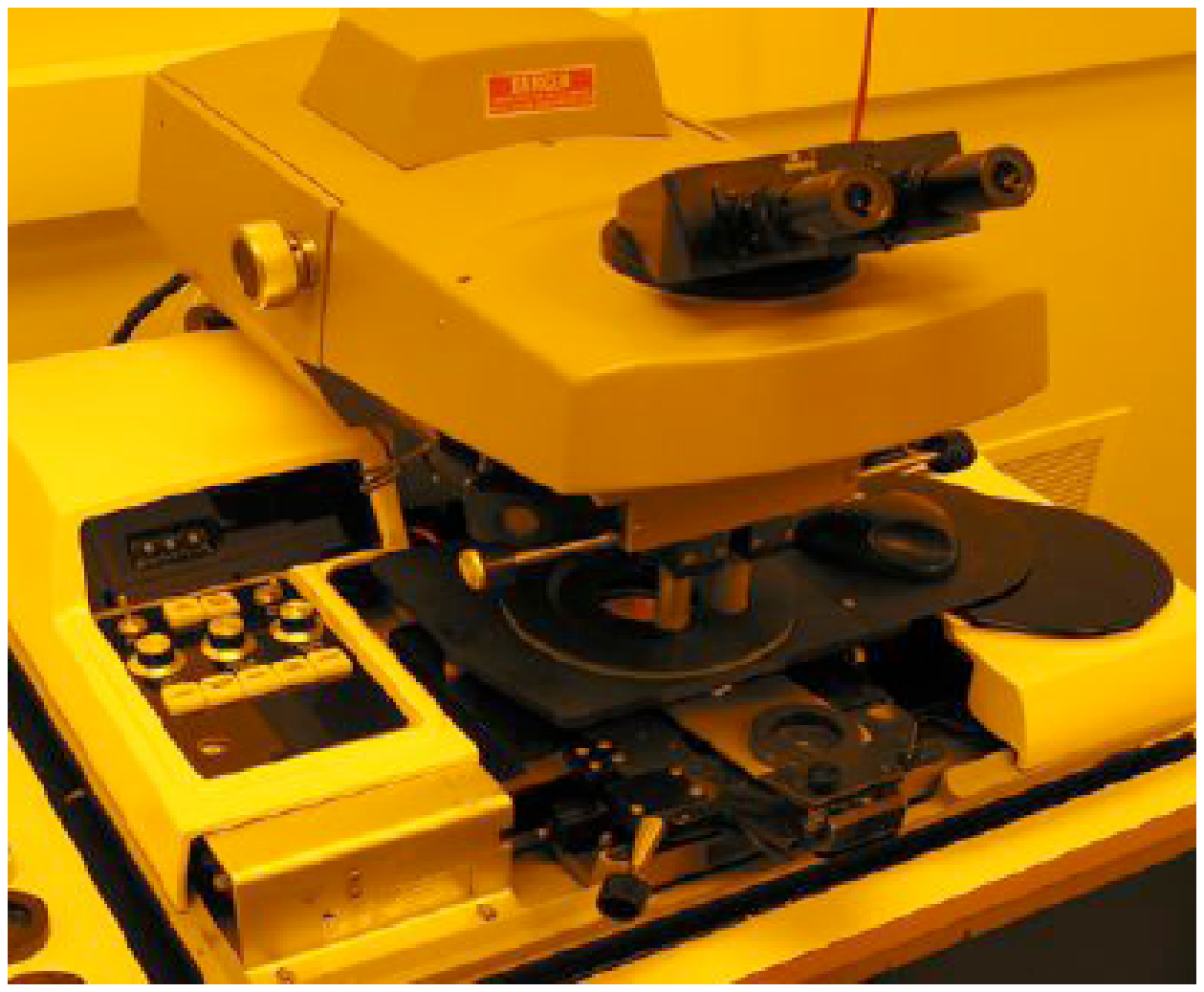}&
    \includegraphics[scale=0.4]{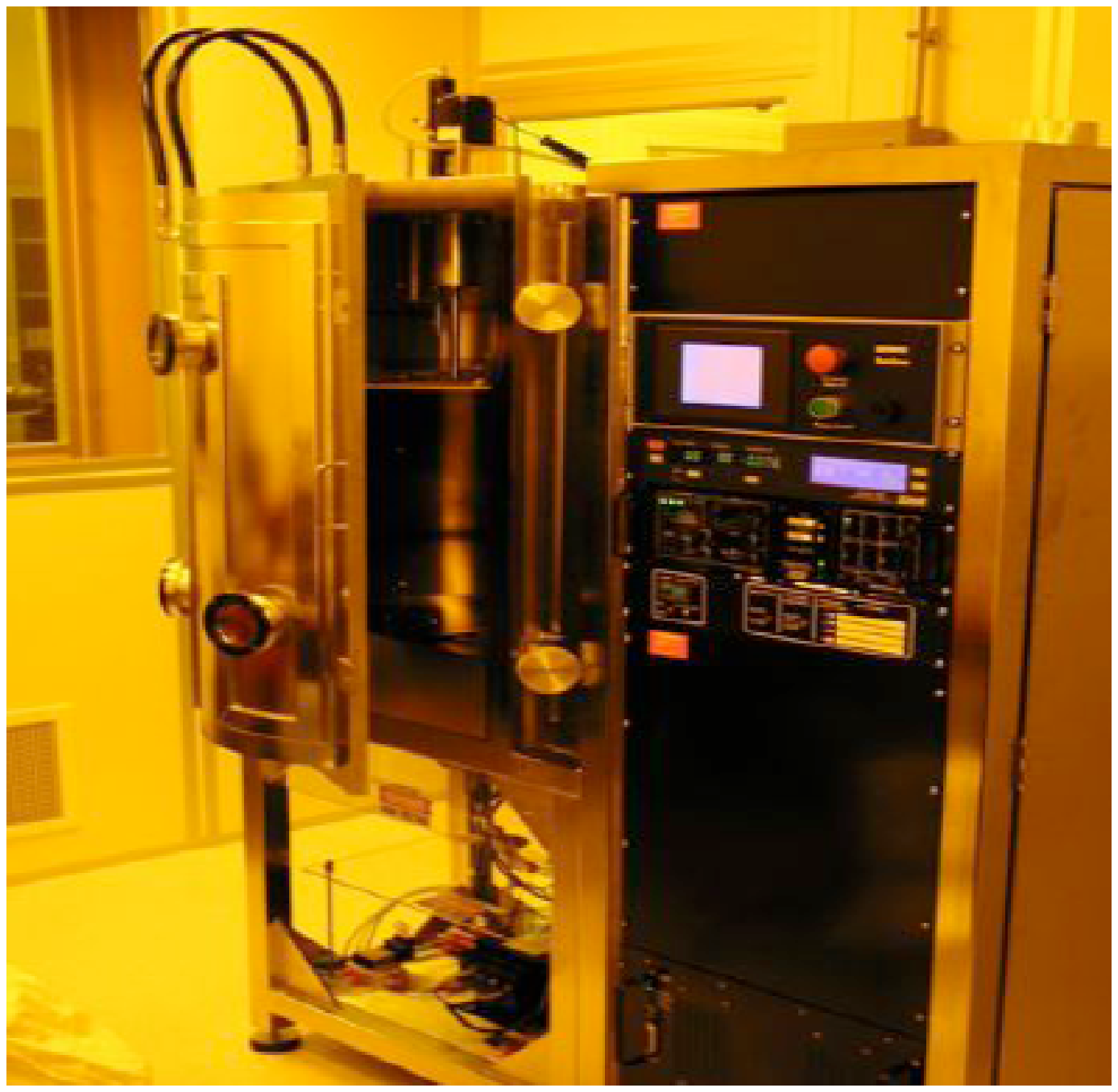}\\
  \end{tabular}
  \end{center}
  \caption{ \label{fig:cleanroom} \small The class-100 clean room operated by our group
  (Physics Department at Washington University in St.Louis). \textbf{Left:} A
  photolithography station to prepare the pixel patterns to be deposited
  onto the detectors. \textbf{Right:} A vacuum electron-beam evaporator to
  deposit thin metal films and/or oxides onto the surface of a crystal.}
\end{figure}

Our semiconductor research group at the Physics Department of Washington
University in St. Louis operates a class-100 clean room (Fig.~\ref{fig:cleanroom}), 
and we work on the development and optimization of pixelated CZT detectors. 
The detectors can be used in X-ray astronomy and in applications related to
medical instrumentation. An example of an application of the CZT detectors 
in astrophysics is the recently developed X-ray scattering polarimeter X-Calibur which is
discussed elsewhere\cite{X-Calibur_SPIE,X-CaliburCalibration}.

This paper is structured as follows: the experimental setup and data analysis
methods are described in Sec.~\ref{sec:setup}, the results on the fine-pitch CZT
detector and CdTe detectors are presented in Sec.~\ref{sec:results},
and the concluding remarks can be found in Sec.~\ref{sec:summary}. 

%%%%%%%%%%%%%%%%%%%%%%%%%%%%%%%%%%%%%%%%%%%%%%%%%%%%%%%%%%%%%
\section{EXPERIMENTAL SETUP AND MEASUREMENTS} 
\label{sec:setup}

The goal of our study is to fabricate and characterise the performance of CZT and CdTe
detectors with pixelated anodes of pixel size $d$ in the range between $100\, \mu$m and 
$250\, \mu$m and pixel pitch $p$ of $200\, \mu$m -- $500\, \mu$m. The detectors are characterized 
by their energy resolution, energy threshold and their detection efficiency as a function 
of energy. We also plan to study the charge sharing between pixels for different pixel pitches.

\subsection{Fine-pitch anode detector fabrication}
\label{sect:setup:cztfab}
\begin{figure}
   \begin{center}
   %\begin{tabular}{cc}
     \includegraphics[scale=1.05]{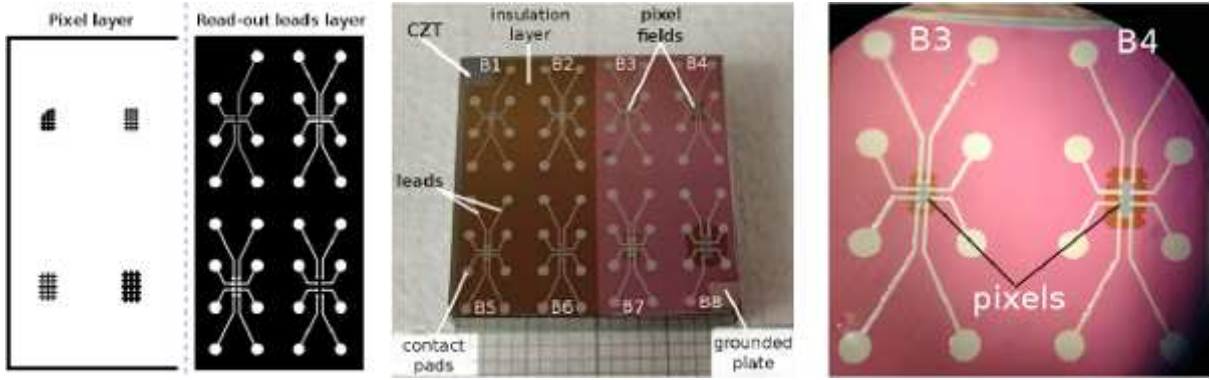}
   %\end{tabular}
   \end{center}
     \caption{ \label{fig:small-mask} \small Photomask layout and fabricated detector with fine-pitch pixels. 
     \textbf{Left:} view of selected layers from the fine-pitch pixel photomask.
     The left side shows the first layer with pixel blocks surrounded by the grounded plate.
     The right side depicts the third layer bearing leads and circular contact pads through
     which the signal is transported from individual pixels.
     \textbf{Middle:} image of the fabricated CZT detector. All eight pixel
     blocks are visible. Pixels and leads/contact pads show up in silver colour. 
     The insulation layer is made of Al$_{2}$O$_{3}$ (brown/pink-coloured material). 
     Different colouring of the insulation layer on the left compared to the right side of 
     the detector stems from the fact that there is no grounded plate deposited on the left 
     side of the detector (for deposition testing purposes).     
     \textbf{Right:} blow-up of the top right quarter of the detector (blocks B3 and B4).
     Clearly visible are pixel blocks consisting of $2 \times 8$ pixels. Pixels are surrounded
     by the grounded plate, whose edge is shaped in a way to mimic a continuation of the pixel 
     grids. Narrow leads (width of $70-100\, \mu$m) carry the signal from pixels to circular contact 
     pads. The contact pads are spaced at a pitch of $2.4\, \rm{mm}$ to match our standard read-out detector 
     fixture capable of reading out $8 \times 8$ pixels (see Sec.~\ref{sec:setup:readout}). 
     Leads/pads are separated from the grounded plate by a layer of insulating Al$_{2}$O$_{3}$ 
     (pink-coloured material).}   
\end{figure}

A photomask with a fine-pitch anode layout was designed and fabricated (Fig.~\ref{fig:small-mask}, 
left). The anode is comprised of eight pixel blocks, each with a different pixel size and pitch 
(see Tab.~\ref{tab:varsize} for values of $d$ and $p$ for each pixel block). Each pixel block
comprises up to eight small pixels. The signal from a pixel is transferred via a lead to a circular 
contact pad that is then connected to the read-out system. In the currently investigated design, 
the anode consists of three layers: 1) \textit{pixel} layer with a grounded plate filling 
the space between the individual pixel blocks, 2) \emph{insulation} layer, and 3) \emph{read-out leads} 
layer. The \emph{pixel} layer is the one deposited directly onto the detector surface, while the \emph{read-out 
leads} layer is the topmost layer (on top of the \emph{insulation} layer) in this sandwich-like design.
In the standard setup of our read-out system, spring-loaded pogo pins (Fig.~\ref{fig:readout})
are used to transfer the signal from detector pixels to ASIC (Application Specific Integrated Circuit) 
channels. However, we are also experimenting with a zebra pad (Fig.~\ref{fig:zpad}, top left) 
and gold wire bonds (Fig.~\ref{fig:zpad}, top right) as alternative methods of signal transfer
from small-sized pixels to ASIC channels. 

The detector's anode was fabricated in our class-100 clean room. The fabrication was
performed in three stages: 1) {pixel} layer, 2) {insulation} layer and 3) {read-out leads}
layer application. Each stage consisted of a photolithography processing followed by 
the deposition of a metal (layer 1 and 3) or an insulator (layer 2) film with an electron 
beam physical vapour deposition system.
The steps in the photolithography process included: pre-baking, applying Microposit S1813 
positive photoresist, post-baking, mask alignment, UV exposure, development in Microposit 
CD-30 Developer, and removing the remaining photoresist with Microposit Remover 1165 after 
a thin film evaporation. The detector surface was polished and etched with Bromine solution 
(5\% Bromine, 95\% Methanol) before carrying out the photolithography and thin film deposition 
steps of stage one. More details on the detector preparation and fabrication can be found 
in Li et al. (2011)\cite{Li2011}. Titanium was used as a material for the pixel/grounded plate layer. 
Aluminium oxide Al$_{2}$O$_{3}$ was used as an insulation material. The fabricated anode on a CZT 
crystal is presented in the middle/right panels of Fig.~\ref{fig:small-mask}. 
The insulation layer of aluminium oxide is $\sim2 \, \rm{k\AA}$ thick, while the pixels/grounded
plate and the leads/contact pads are $\sim 1.25\, \rm{k\AA}$ and $\sim 3\, \rm{k\AA}$ thick, respectively. 
The CZT detector used in our study has a monolithic gold cathode. The design/fabrication of multiple 
pixel blocks has the advantage to simultaneously test different pixel geometries on the same detector 
in one fabrication cycle.
\begin{table}
    \begin{center}
	\caption{Parameters of pixel blocks B of the fine-pitch pixelated anode (see Fig.~\ref{fig:small-mask} 
	for the positions of the blocks B1 to B8). Along the values of pixel size $d$ and pixel pitch $p$, the number 
	of pixels in each block is given.}
	\begin{tabular}{cccc}
	\multicolumn{4}{c}{}\\ \hline
	Block & $d$ & $p$ & No. \\ 
	name & [$\mu$m] & [$\mu$m] & pixels \\ \hline \hline
	B1  & 100 & 250 & 7 \\
	B2  & 150 & 250 & 8 \\
	B3  & 100 & 200 & 8 \\
	B4  & 100 & 300 & 8 \\
	B5  & 200 & 300 & 8 \\
	B6  & 150 & 350 & 8 \\
	B7  & 250 & 350 & 8 \\
	B8  & 150 & 500 & 7 \\ \hline
	\end{tabular}
    \end{center}
\label{tab:varsize}
\end{table}

\subsection{Read-out system}
\label{sec:setup:readout}

   \begin{figure}
   \begin{center}
   \begin{tabular}{c}
     \includegraphics[scale=0.63]{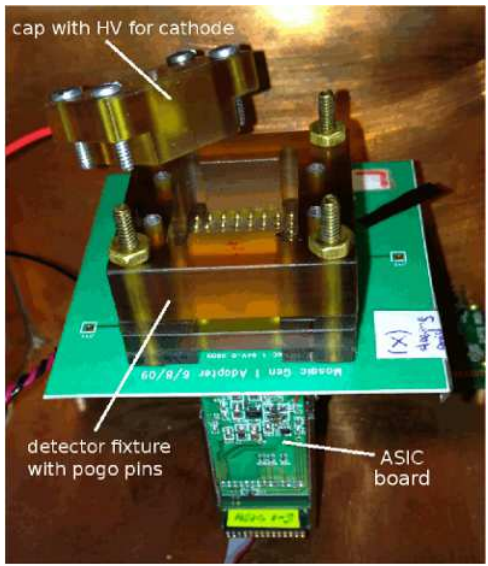}
   \end{tabular}
   \end{center}
     \caption{ \label{fig:readout} The detector read-out system. Shown here is the adapter
     board (designed at Washington University) with the detector fixture mounted on top. 
     The two BNL 32-channel ASICs are plugged into the bottom
     side of the board. The detector is kept in the fixture by the cap through which the high 
     voltage is supplied to the cathode. The detector's anode side presses onto spring-loaded pogo 
     pins making contact to the pads (see Fig.~\ref{fig:small-mask}). 
     The X-ray source (not shown in the figure) is placed centred above the detector's cap.}
   \end{figure}

Figure \ref{fig:readout} shows the read-out system used to collect the signal from the
fine-pitch detector. The detector is placed in the Ultem plastic fixture that is mounted 
on top of the adapter board. The gold-coated, spring-loaded pogo pins connect pixels
with traces on a printed circuit board. The detector is kept in place with an Ultem
cap. The cap is equipped with one pogo pin that is utilized to provide the high voltage
to the cathode of the detector. The detector is read out by two 32-channel ASICs developed by G. De Geronimo 
(Brookhaven National Laboratory, BNL) and E. Wulf (Naval Research Laboratory, NRL) \cite{wulf2007}. 
The adapter boards were designed at Washington University. Each channel is triggered by an adjustable 
threshold discriminator in the ASIC. The ASICs have a readout noise of around $2.5\, \rm{keV}$ FWHM.
The ASIC channels were operated at a medium gain setting of $28.5 \, \rm{mV/fC}$ and a
signal shaping time of $0.5\, \mu$s.

The fine-pitch anode mask was designed in such a way that it can be read out by our
current read-out system without the necessity of bump bonding the detector to the adapter
board. This option offers a freedom in testing different anode designs simply by re-fabricating
the detector surface. In a later step a detector with a full-sized fine-pitch 
anode will be fabricated and bump bonded for follow-up tests with a low noise small 
pixel ASIC (see Sec.~\ref{sec:setup:lownoise}).

   \begin{figure}
   \begin{center}
   \begin{tabular}{cc}
        \includegraphics[scale=0.70]{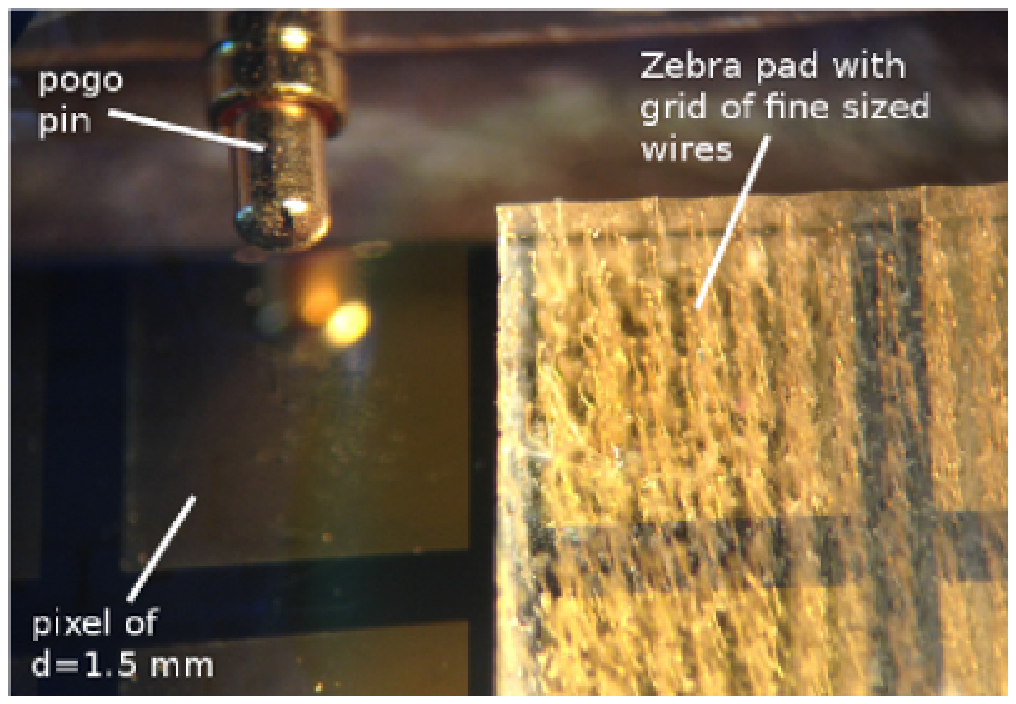}&
        \includegraphics[scale=0.37]{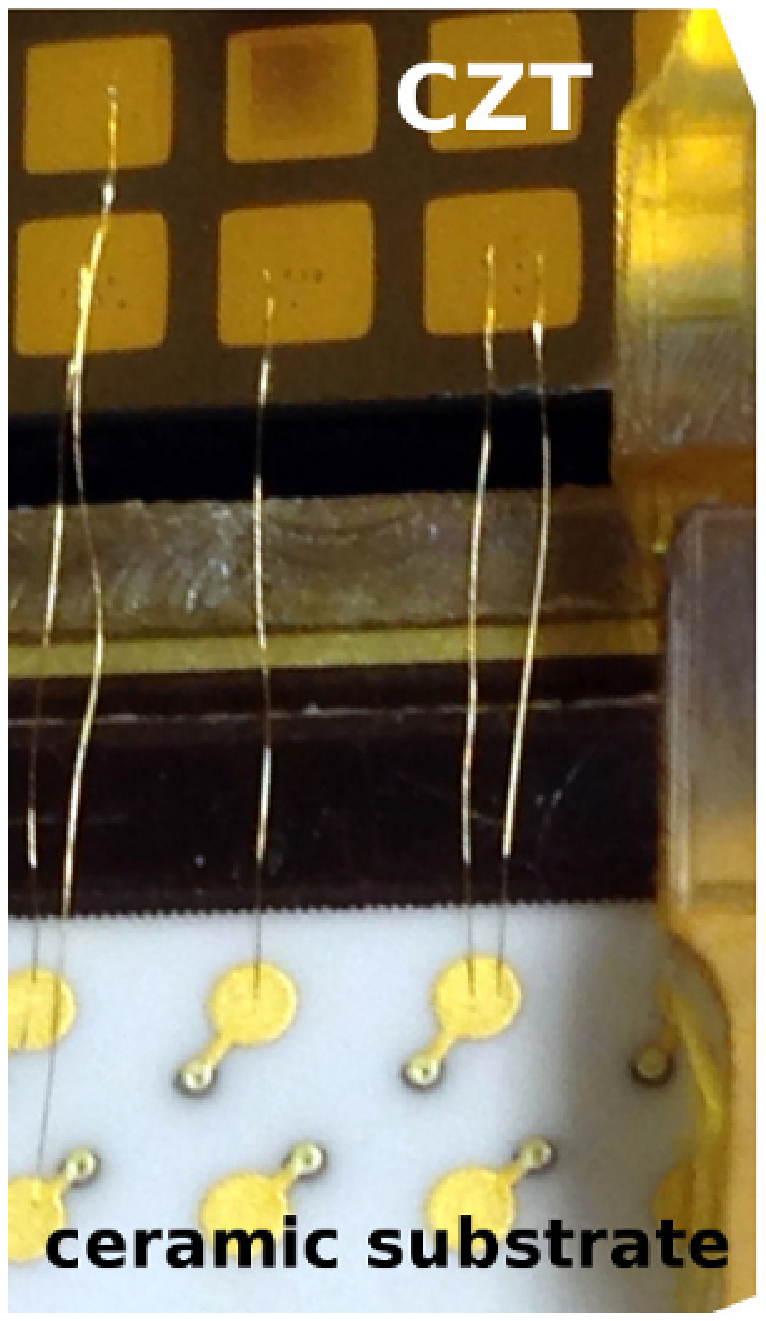}\\
        \includegraphics[scale=0.4]{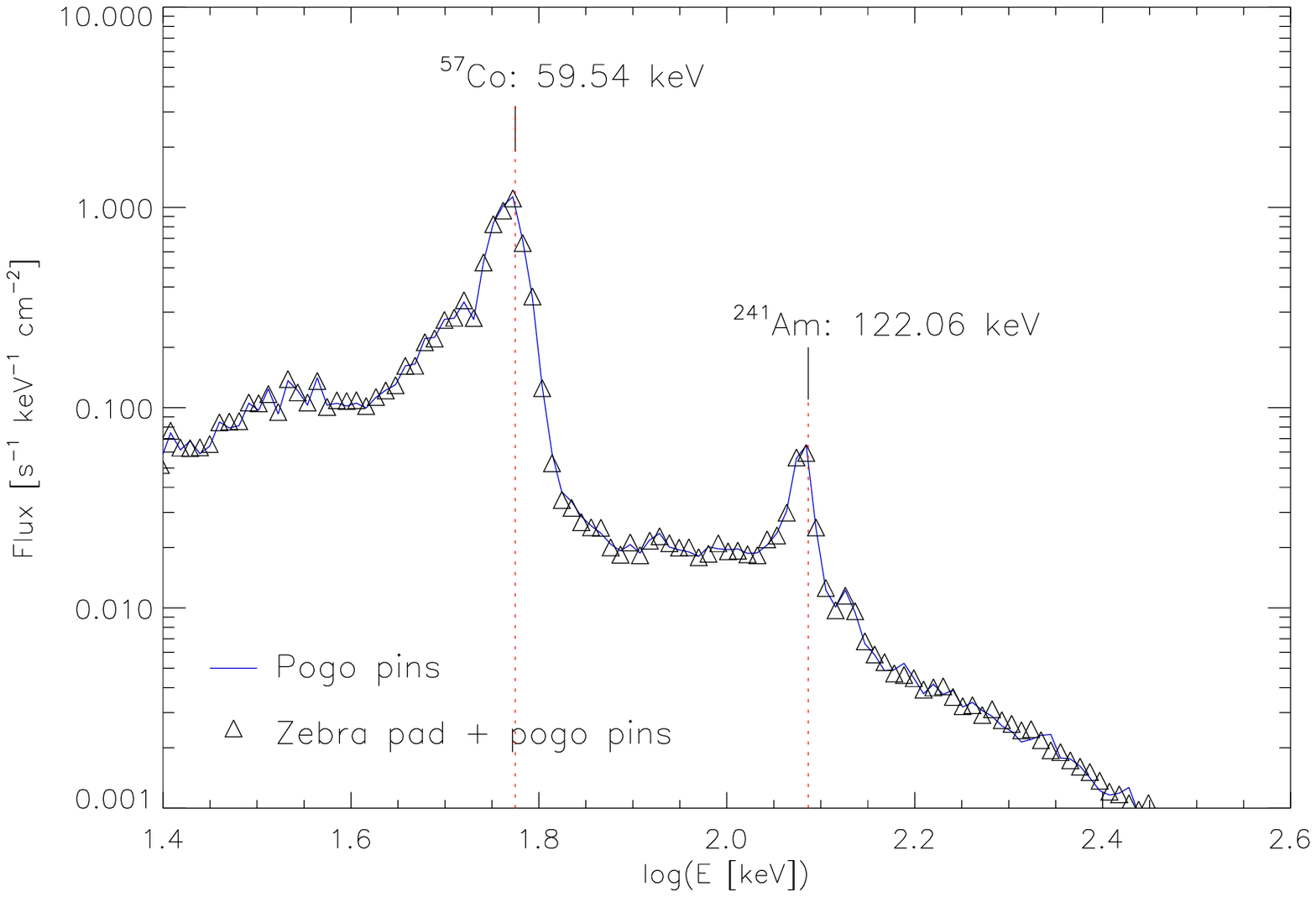}&
        \includegraphics[scale=0.35]{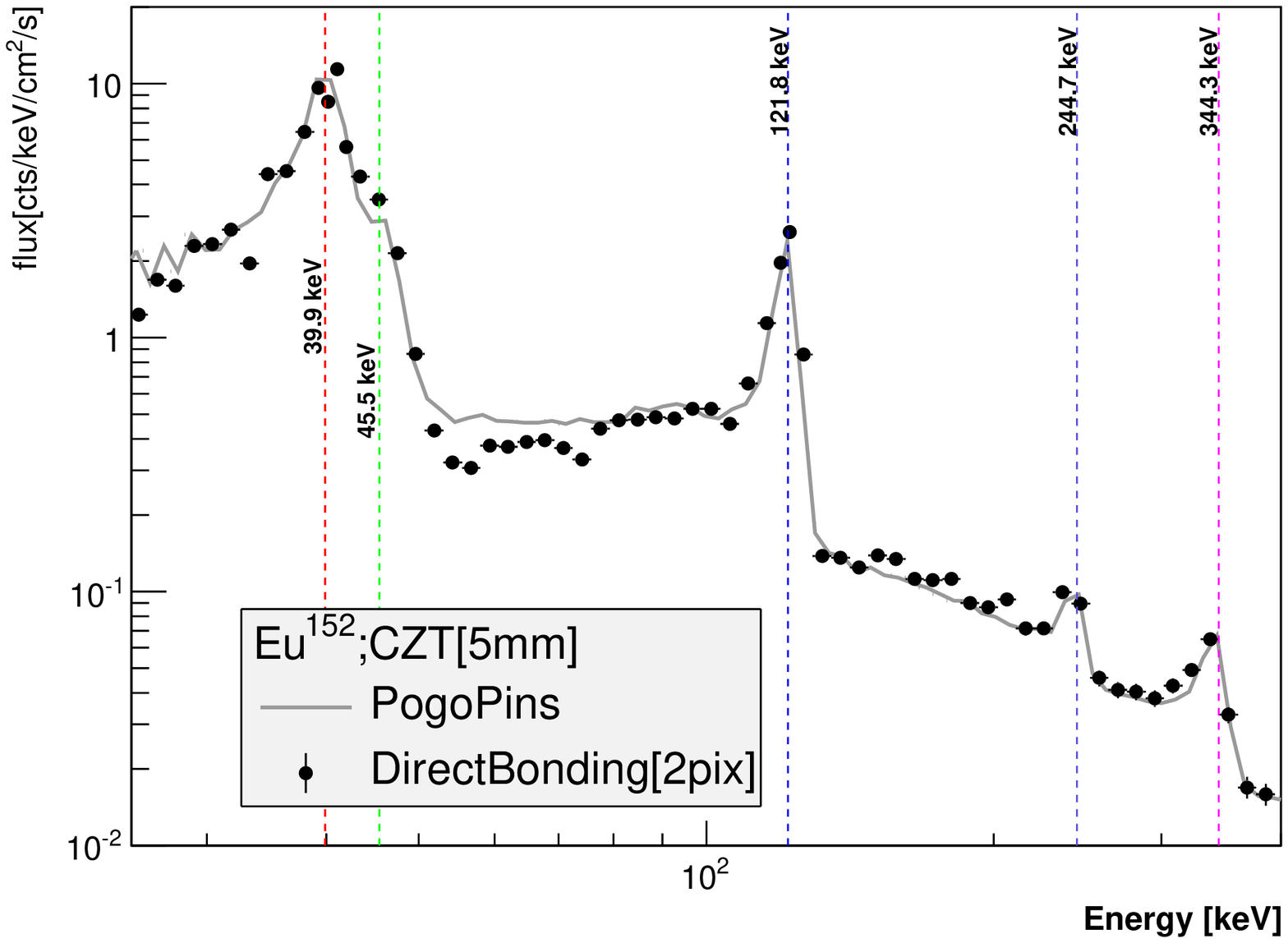}\\
   \end{tabular}
   \end{center}
     \caption{ \label{fig:zpad} Different methods for transfer of signal from the detector's anodes
     to the read-out system: pixel/zebra pad/pogo pin combination (\textbf{top left}) and direct 
     wire bonding between pixels and a ceramic adapter board (\textbf{top right}), which in turn is
     plugged into the ASIC board. 
     Spectra obtained with each of the tested methods are presented in the bottom row. 
     The \textbf{bottom left} panel shows data obtained with the system where the zebra pad was
     inserted between the detector and the pogo pins, while the \textbf{bottom right} panel shows measurements
     obtained from two wire-bonded pixels. In each case, the spectra are contrasted with the average 
     detector spectrum obtained with our standard read-out method, i.e. pogo pins. $^{57}$Co, $^{152}$Eu 
     and $^{241}$Am radioactive sources were used in the tests. The presented tests of different options
     for signal transfer were carried out with a regular size 2.4 mm pixel pitch detector.}
   \end{figure}
   
The signal-transfer methods, in addition to the standard pogo pins, that are currently 
under consideration are zebra pad\footnote{Zebra pad is a silicon pad with a grid of fine-sized
wires embedded into the pad. An example of zebra pad is shown in Fig.~\ref{fig:zpad} (top left panel). 
By using a zebra pad we want to reduce the localized impact of pogo pins on the CZT
crystal, and also on the thin metal/insulator layers of the sandwich-like structure 
of the small-sized anodes (i.e. prevent punching through the thin anode layers and shortening
the contact pads to the grounded plate).} placed between the pogo pins and the anode pads 
on the detector and direct gold wire bonds between the pixels and an ASIC/channel 
adapter board. The zebra pad would reduce the localized impact on the CZT crystal caused 
by a pogo pin point of contact. The direct wire bonding to pixels would simplify the detector 
fabrication and lead to lower resistance of the contact leads - if it can be applied 
successfully to fine pixel pitches. The two methods were tested on a detector 
purchased from Quikpak/Redlen\footnote{\tt http://redlen.ca} with an anode of $8 \times 8$ 
pixels at $2.4\, \rm{mm}$ pitch and are illustrated in Fig.~\ref{fig:zpad} (top).
In the bottom row of Fig.~\ref{fig:zpad} spectra obtained with each of the tested methods are presented.
The energy calibrated spectra are also contrasted with the reference spectrum that
was obtained with the system where only pogo pins were used to transfer the signal from
pixels to ASIC channels. Though both, the zebra pad and the wire bonds, extend
the travel path of the signal potentially exposing it to a pick up noise, no deterioration
of spectral resolution and no increase of the noise component was observed in 
the obtained spectra. Thus, both signal-transfer methods can be used at the current stage 
of our project without any spurious effects being introduced into the collected data.
The application of both approaches to pixels with pitches of $200\, \mu$m is currently being tested.

\subsection{Development of a low noise, small pixel ASIC}
\label{sec:setup:lownoise}

To reduce the readout noise and to optimize the energy resolution and
energy threshold of the individual pixels, it is advantageous to directly
bond the readout ASIC to the detector pixels - with the ASIC/channel contact 
pads matched to the pixels of the detector. A number of ASICs with 2D pad arrays 
have been developed for CZT detectors with $350 - 600\, \mu \rm{m}$ pixel pitches. 
Our programme includes the design of the complete front-end channel 
of a 2D-ASIC (control logic and the physical layout), following the approach of 
ASIC designs for smaller-pixel-pitch detectors existing for Si-detectors.

A prototype front-end ASIC has been designed (Brookhaven National
Laboratory) and fabricated in a commercial 130-nm process. The ASIC has
a 16-by-16 channel array at the pitch of $250 \, \mu\rm{m}$. Each
channel includes a low-noise charge amplifier with adjustable gain, a
shaper with baseline stabilizer at adjustable time constant (of peaking
time 125, 250, 500, and $1000\, \rm{ns}$), a discriminator, a peak-detector, and
a 6-bit address register. Shared among channels are a pulse generator, 
bias circuits, global registers, and a low-voltage differential 
interface. Modes of operation are the following: acquisition, when events are detected
and processed; readout, when events are read out in sparse mode with
token passing; and configuration, when global and channel registers are
accessible for configuration.

First tests of the ASIC are currently being performed. We plan to test the
prototype of the newly developed ASIC next year, once the ASIC tests are
finalized and the ASIC is directly bonded to a detector with a matching
pixel grid. These developments will address the specific issues important
for the use of the ASIC in X-ray astronomy: a low energy threshold and
an excellent energy resolution.

\subsection{Data analysis}
\label{sec:data:analysis}
Data presented in this paper were taken with either a $^{152}$Eu radioactive source or the radioactive
source combination of $^{57}$Co and $^{241}$Am.
The $^{152}$Eu source, with a half-life of 13.5 years, has emission lines at $39.8\, \rm{keV}$,
$121.8\, \rm{keV}$, $244.7\, \rm{keV}$ and $344.3\, \rm{keV}$. The $^{57}$Co source, with a half-life of 271.8 days,
shows emission lines at $14.4\, \rm{keV}$, $122.1\, \rm{keV}$ and $136.5\, \rm{keV}$. The $^{241}$Am radioisotope
has a half-life of 432.6 years and emission lines at $13.6\, \rm{keV}$, $16.9\, \rm{keV}$ and 
$59.5\, \rm{keV}$. During measurements the radioactive source was suspended on the cathode side approximately $1-2\, \rm{cm}$ 
above the detector's geometrical centre.

\textit{Energy calibration:} Using a radioactive source placed above the detector, the energy 
scale of the individual pixels/channels was calibrated. In this step the fitted peak position 
of the emission lines in the raw uncalibrated spectrum is compared to the nominal energy of 
the selected lines in order to obtain the pedestal and amplification of each active ASIC 
channel. In the case where the detector was illuminated with $^{152}$Eu, lines at $39.8\, \rm{keV}$ 
and $121.8\, \rm{keV}$ were used for calibration. The $59.5\, \rm{keV}$ line of $^{241}$Am and
$122.1\, \rm{keV}$ line of $^{57}$Co were used for calibration when $^{241}$Am and $^{57}$Co
were used in the measurements.

\textit{Characterisation of emission lines:} The emission lines in the energy calibrated
spectra were fitted with an asymmetric Gaussian function. The fitted profile allows for 
offsets and asymmetric widths $\sigma_{1}$ and $\sigma_{2}$ on the left and on the right
side of the peak position in order to account for possible asymmetries in the line shapes.
Whenever the $^{152}$Eu source was used in the measurements, its low energy line at $39.8\, \rm{keV}$
was fitted with a profile consisting of a sum of two Gaussians. This fit accounts 
for the existence of the lower amplitude emission line at $45.7\, \rm{keV}$, which at a few keV energy
resolution is blended with the $39.8\, \rm{keV}$ line. Thus, if not taken into account, this line may cause 
systematic shifts in the fitted peak positions. The energy resolution is defined as 
the full width at half maximum FWHM of the fitted asymmetrical Gaussian function 
$\Delta E = \rm{FWHM} = 2\sqrt{2\ln2} \cdot \frac{1}{2} (\sigma_{1}+\sigma_{2})$.
In the presented work the energy resolution is either given in units of keV or in relative
units of $\Delta E / E$ (here $E$ is the energy of the fitted emission line).

\section{RESULTS}
\label{sec:results}

% ---------------------- TESTED DETECTORS: internal use -----------------------------------------
% ID of CZT and CdTe detectors used in the measurements:
% CZT 4273 -> fine-pitch pixelated anode tests
% CdTe 1303-1301-4 -> thin 1x1cm2 CdTe detector with small 8x8 pixels+guard ring: temp. and HV tests
% CdTe 1210-2202-1 -> thin 2x2cm2 CdTe detector with regular 8x8 pixels: I/V, temp., HV tests
% --> see wiki page on 2x2cm2 CdTe fabrication and tests
% -----------------------------------------------------------------------------------------------

\subsection{Characterisation of CZT detectors with fine-pitch pixelated anode}
\label{sec:results:czt}
A fine-pitch pixelated anode was successfully fabricated on a $20 \times 20 \times 5\, \rm{mm}^{3}$ 
CZT detector (Fig.~\ref{fig:small-mask}) purchased from Quikpak/Redlen. 
The detector was inserted into the existing read-out system, and the data are currently being taken. The $^{152}$Eu 
radionuclide is used as a source of X-rays. The progress on the characterisation of the CZT detector with
the fine-pitch pixelated anode will be reported at the SPIE 2014 conference.

%   \begin{figure}[t]
%   \begin{center}
%   %\begin{tabular}{c}
%        \includegraphics[scale=0.4]{images/PreliminarySpectrum_B1B2_Multi1.eps}
%   %\end{tabular}
%   \end{center}
%     \caption{ \label{fig:b1spec} Preliminary spectra of $^{152}$Eu source taken with a CZT detector
%     with a fine-pitch anode (see Sec.~\ref{sect:setup:cztfab}). The presented spectra are pixel-averaged 
%     spectra of selected pixel blocks: B1 (two pixels active) and B2 (two pixels active). The two pixel blocks 
%     B1 and B2 are not surrounded by the grounded plate.}
%   \end{figure}

\subsection{Characterisation of thin CdTe detectors} 

\begin{figure}
\begin{center}
 \begin{tabular}{cc}
   \includegraphics[scale=0.36]{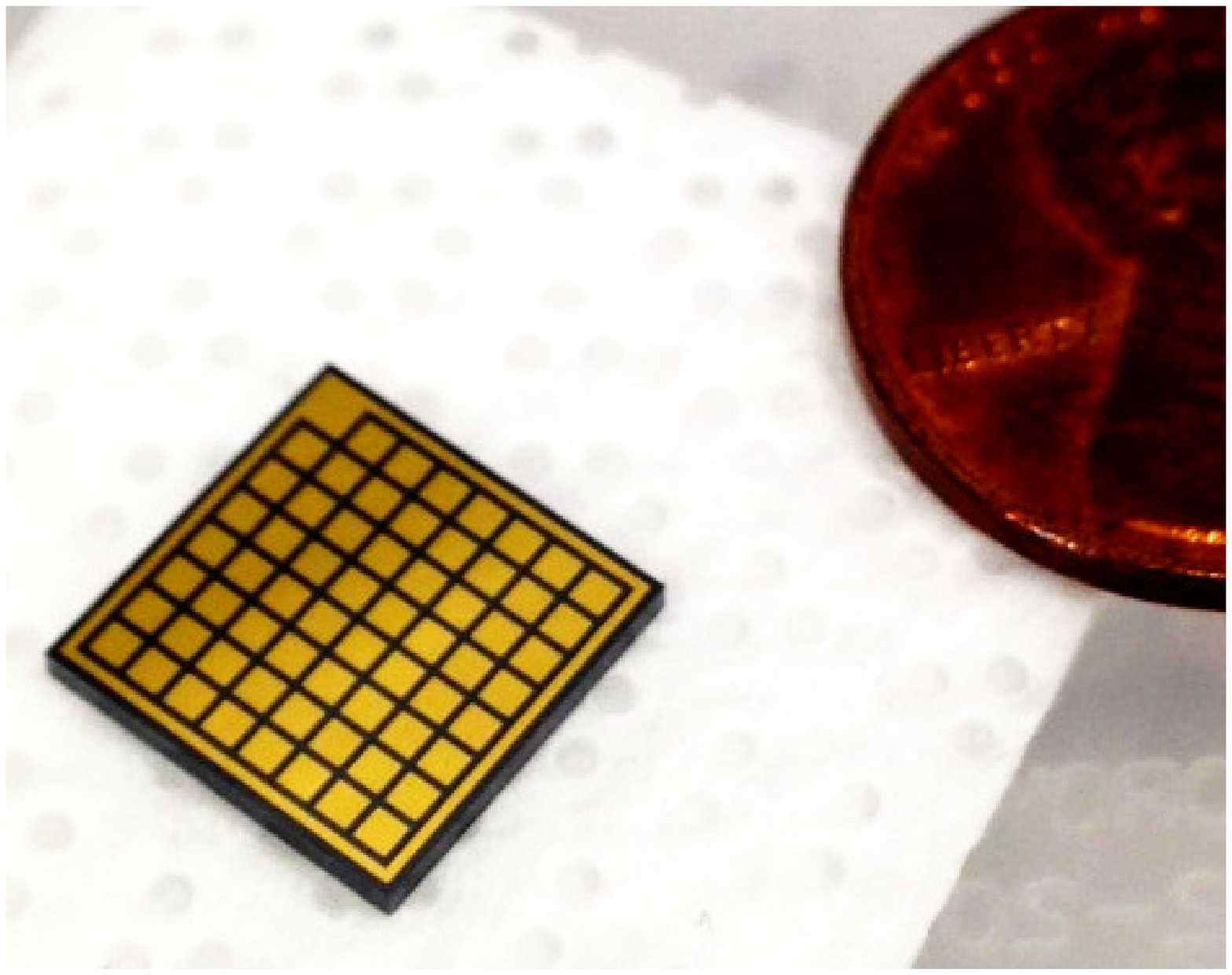}&
   \includegraphics[scale=0.39]{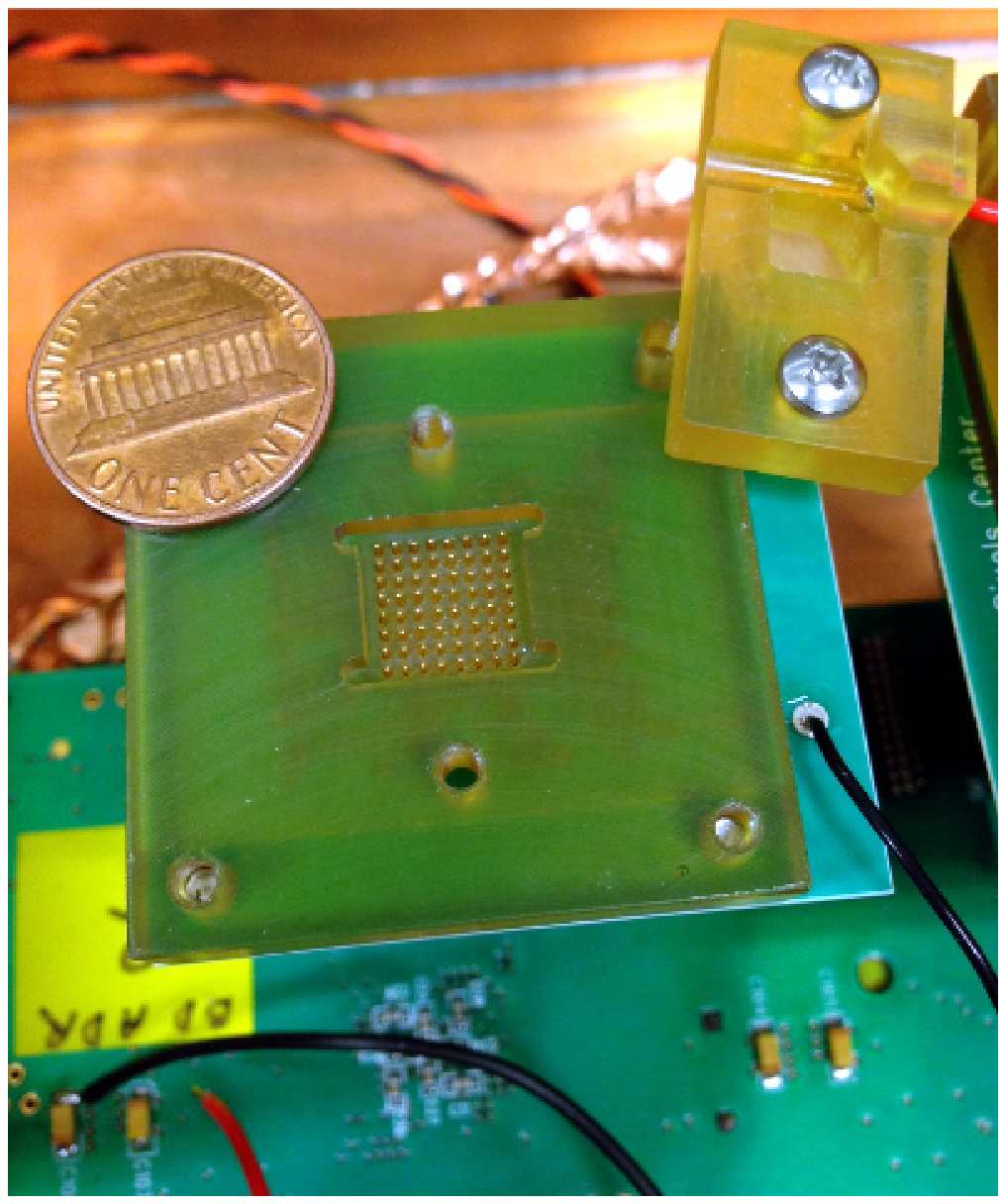}\\
 \end{tabular}
 \end{center}
 \caption{\label{fig:Acrorad_CdTe}Acrorad CdTe detectors ($10 \times 10
 \times 1 \, \rm{mm}^{3}$, pixel pitch $1.1 \, \rm{mm}$, pad size $0.88
 \times 0.88 \, \rm{mm}^{2}$) with Pt cathode and Al/Ti/Au/Ni/Au pixels.
 \textbf{Left:} One of the five detectors as bought from Acrorad. 
 \textbf{Right:} Ultem fixture and HV (designed and fabricated at Washington
 University), connected to the X-Calibur read-out system\cite{X-Calibur_SPIE,X-CaliburCalibration}.}
\end{figure}

\begin{figure}
  \begin{center}
  \begin{tabular}{ccc}
      \hspace{-0.35cm}\includegraphics[scale=0.28]{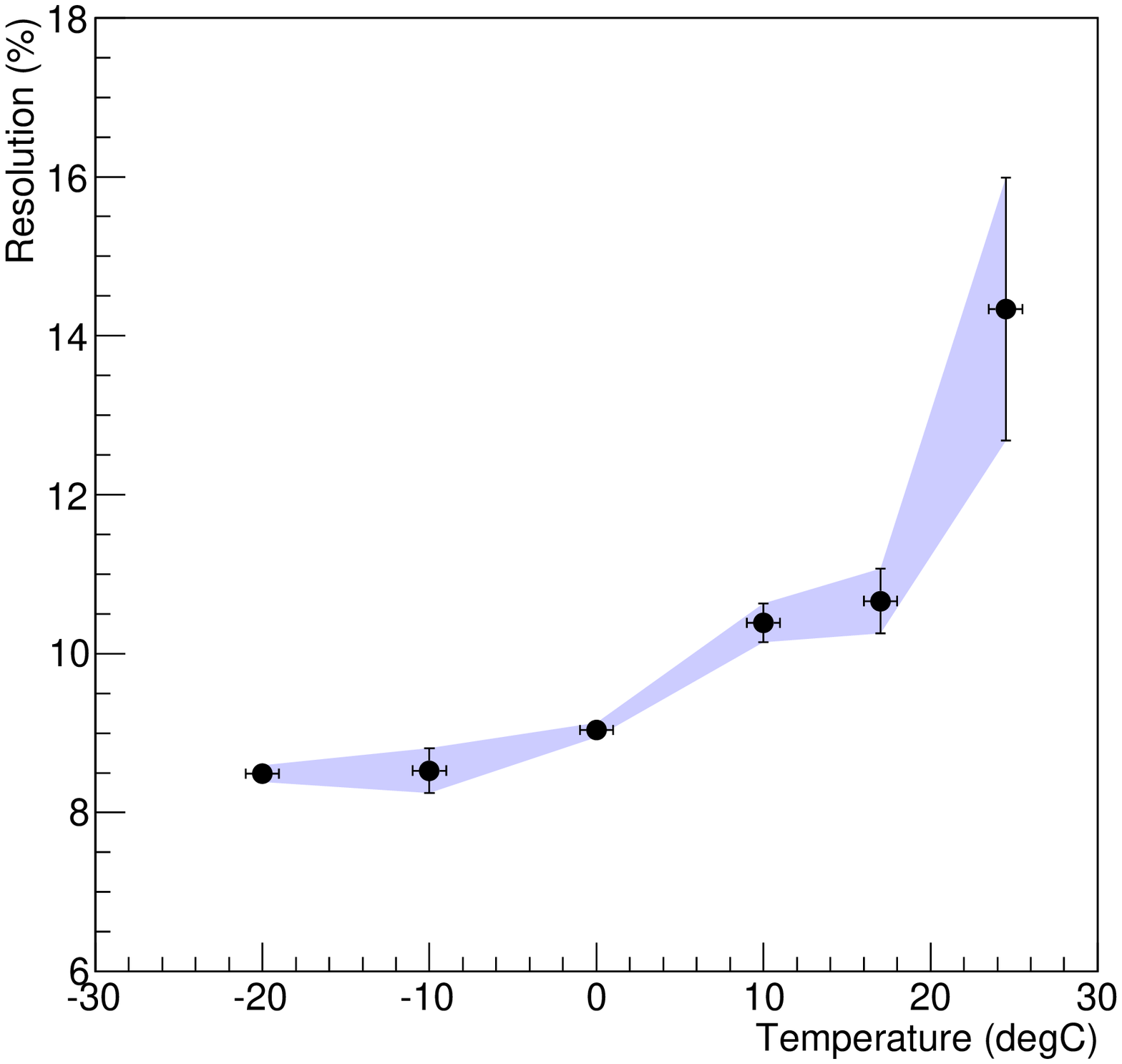}&
      \hspace{-0.7cm}\includegraphics[scale=0.28]{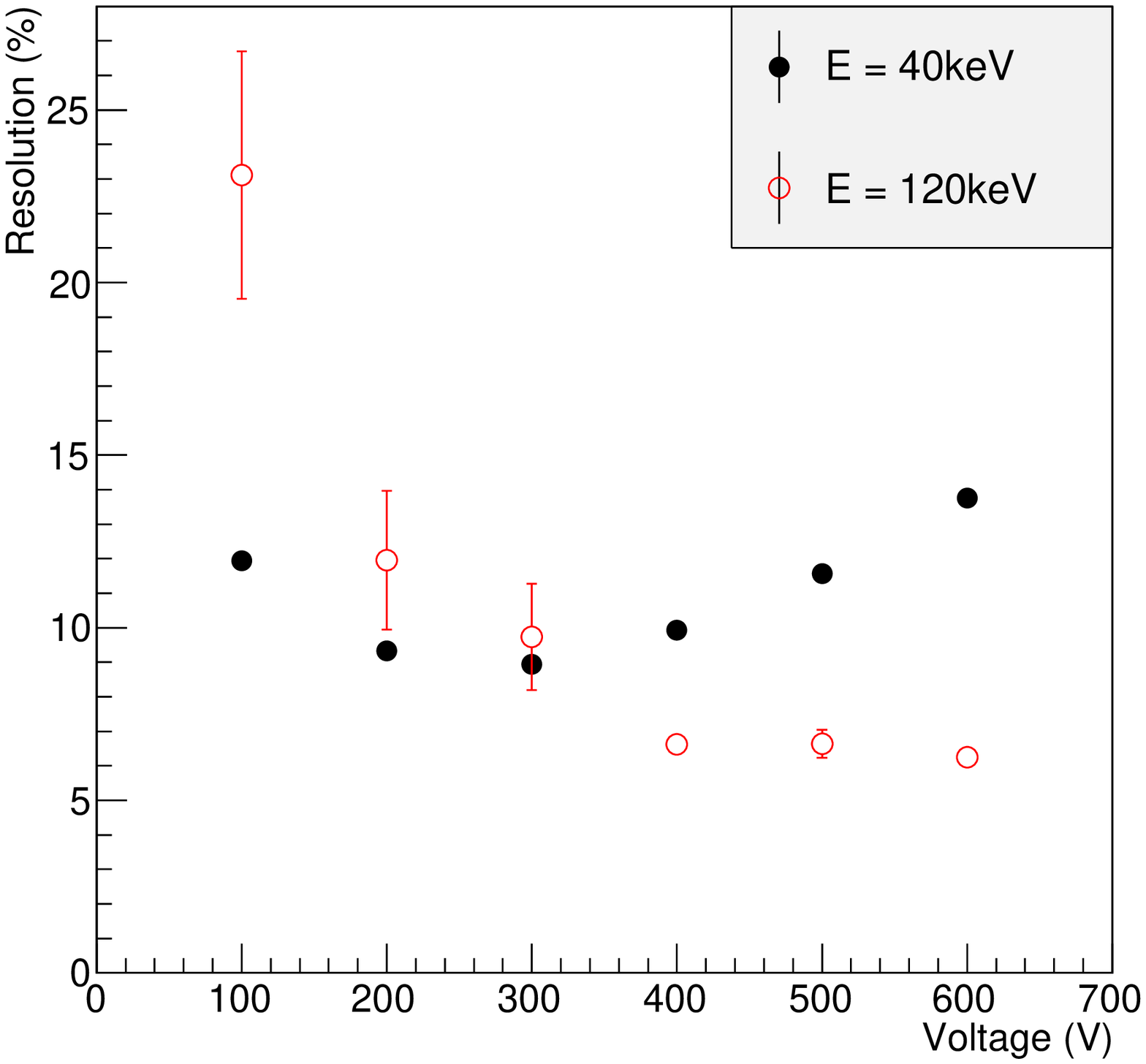}&
      \hspace{-0.58cm}\includegraphics[scale=0.33]{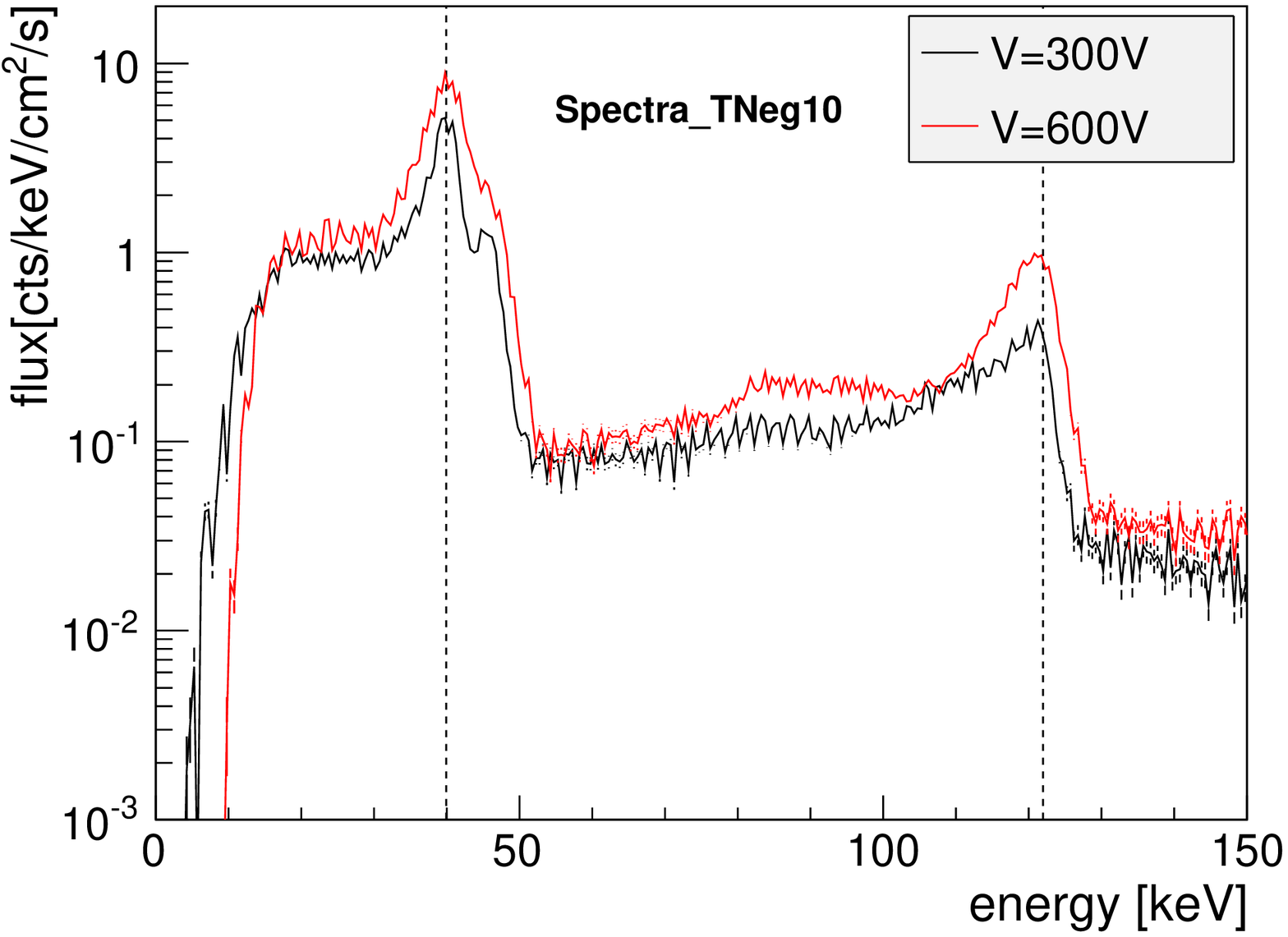}\\
  \end{tabular}
      \caption{ \label{fig:cdtetests} \textbf{Left:} energy resolution of the $40\, \rm{keV}$ 
      energy line of $^{152}$Eu as a function of temperature and as a function of time. The lower line 
      of the band characterises the resolution after 10 minutes of data had been taken, while the upper 
      line shows the resolution after 30~min (for $T = 10^{\circ}$C and 17$^{\circ}$C) or 50~min (for 
      all other $T$). Note, however, that the detectors had been biased for $30-60$ minutes prior to the start 
      of the measurements (threshold optimization).
      \textbf{Middle:} resolution of the $40\, \rm{keV}$ and $120\, \rm{keV}$ energy lines of $^{152}$Eu
      as a function of bias voltage. The data points result from 80-minute data runs.
      \textbf{Right:} energy spectra of $^{152}$Eu measured at two different voltages 
      at $T = -10^{\circ}$C.}
  \end{center}
\end{figure}

Most detector backgrounds in space-missions scale with the volume of the detector crystal. 
This scaling leads to the choice of smaller detector thicknesses without compromising the X-ray absorption 
capabilities for $E \leq 100 \, \rm{keV}$. We therefore expanded our detector repository 
with five pixellated $10 \times 10 \times 1 \, \rm{mm}^{3}$ CdTe detectors (Fig.~\ref{fig:cdtetests}, left) bought
from Acrorad\footnote{\tt http://www.acrorad.co.jp/us/}. 
We designed and fabricated a pogo pin fixture that matches the detectors with 
the smallest footprint and enables us to read them with our existing read-out infrastructure 
(Fig.~\ref{fig:Acrorad_CdTe}, right). The CdTe detectors have monolithic platinum cathode. 
Their pixelated anode is made of Al/Ti/Au/Ni/Au thin metal layers and is comprised of $8 \times 8$ pixels 
with $1.1\, \rm{mm}$ pitch. The pixel grid is surrounded by a guard ring, which was set to ground throughout 
the measurements. All tests were performed with $^{152}$Eu as the X-ray source with the detector/read-out
system placed inside a temperature chamber.

In contrast to CZT detectors, CdTe detectors show an effect of increasing dark current with time. 
First, the steady growth of dark current with time is observed, then this growth is followed by an instantaneous 
increase in dark current level by an order of magnitude. This effect is referred to as a breakdown 
phenomenon\cite{Garson2005}. At higher temperatures, the dark current builds up more quickly until the point 
after which reliable measurements are no longer possible (in our case, starting after $\sim$50 minutes 
at room temperature and after $\sim$2.5 hours at $-10^{\circ}$C). Therefore, operation at lower temperatures 
will guarantee a more stable and lengthier operation of the detectors. 

\textit{Temperature tests of CdTe detectors:} 
To test how the detector behaved at different temperatures, we analysed data from the detector biased to 
$-300\, \rm{V}$ and operated at various temperatures (25$^{\circ}$C, 17$^{\circ}$C, 10$^{\circ}$C, 0$^{\circ}$C, $-10^{\circ}$C, 
and $-20^{\circ}$C). The trend of the energy resolution of the $40\, \rm{keV}$ peak for $^{152}$Eu is shown 
in Fig.~\ref{fig:cdtetests} (left). The boundaries of the band indicate the drift of resolution of the $40\, \rm{keV}$ 
peak with time. 
As displayed by the graph, the energy resolution worsens not only as the CdTe detector is at higher 
temperatures but also as this detector is working for longer\footnote{Degradation of energy resolution 
with operation time observed for CdTe detectors is caused by a polarization effect\cite{Bell1974, Cola2009}. 
When a bias voltage is applied and photon interaction occurs in the CdTe crystal, liberated electrons and holes 
travel towards their respective electrodes. In the presence of trapping sites, some of the charges can be 
trapped resulting in a reduction/modification of the internal electric field experienced by the carriers. 
The decrease of the electric field reduces drift velocities of the charges and in turn more carriers 
are being trapped. This polarization effect causes a decrease in counting rate and pulse amplitude with time. 
It is important to mention that this effect is most severe when operating at high temperatures and low bias 
voltages.}. For each temperature, we optimized the ASIC channel trigger thresholds, which we found we must 
increase more at temperatures lower than $-10^{\circ}$C (the thresholds were found to be around $15\, \rm{keV}$). 
From this factor, in combination with the resolution graph, we concluded that of the temperatures we tested, 
$-10^{\circ}$C is the optimal temperature at which to run the CdTe detector.  

\textit{Bias voltage dependence of energy resolution for CdTe:}
Given that we determined $-10^{\circ}$C to be the optimal temperature at which to use the CdTe detector, 
we kept our system in this environment as we tested different bias voltages ($-50\, \rm{V}$, $-100\, \rm{V}$,
 $-200\, \rm{V}$, $-300\,  \rm{V}$, $-400\, \rm{V}$, $-500\, \rm{V}$, and $-600\, \rm{V}$).  
As Figure ~\ref{fig:cdtetests} (middle) displays, the trends of energy resolution vs. bias voltage differ 
according to the peak studied. 
For the $40\, \rm{keV}$ peak of $^{152}$Eu, $-300\, \rm{V}$ is the optimum for resolution. 
At $-50\, \rm{V}$, the $40\, \rm{keV}$ line is no longer well-defined (merging with the threshold regime). 
For the $120\, \rm{keV}$ peak, however, the resolution improves at higher voltages, beginning to level off 
after $-500\, \rm{V}$.  Figure~\ref{fig:cdtetests} (right) shows this difference in resolution at $-300\, \rm{V}$ 
and $-600\, \rm{V}$. The $-600\, \rm{V}$ provides better resolution at the high energy lines.
We are currently testing 4 other $10 \times 10 \times 1\, \rm{mm}^{3}$ CdTe detectors. 
These detectors will also be used in the long-term to test sub-mm pixel pitches.

\begin{figure}
\begin{tabular}{cc}
%\centering{
\includegraphics[scale=0.4]{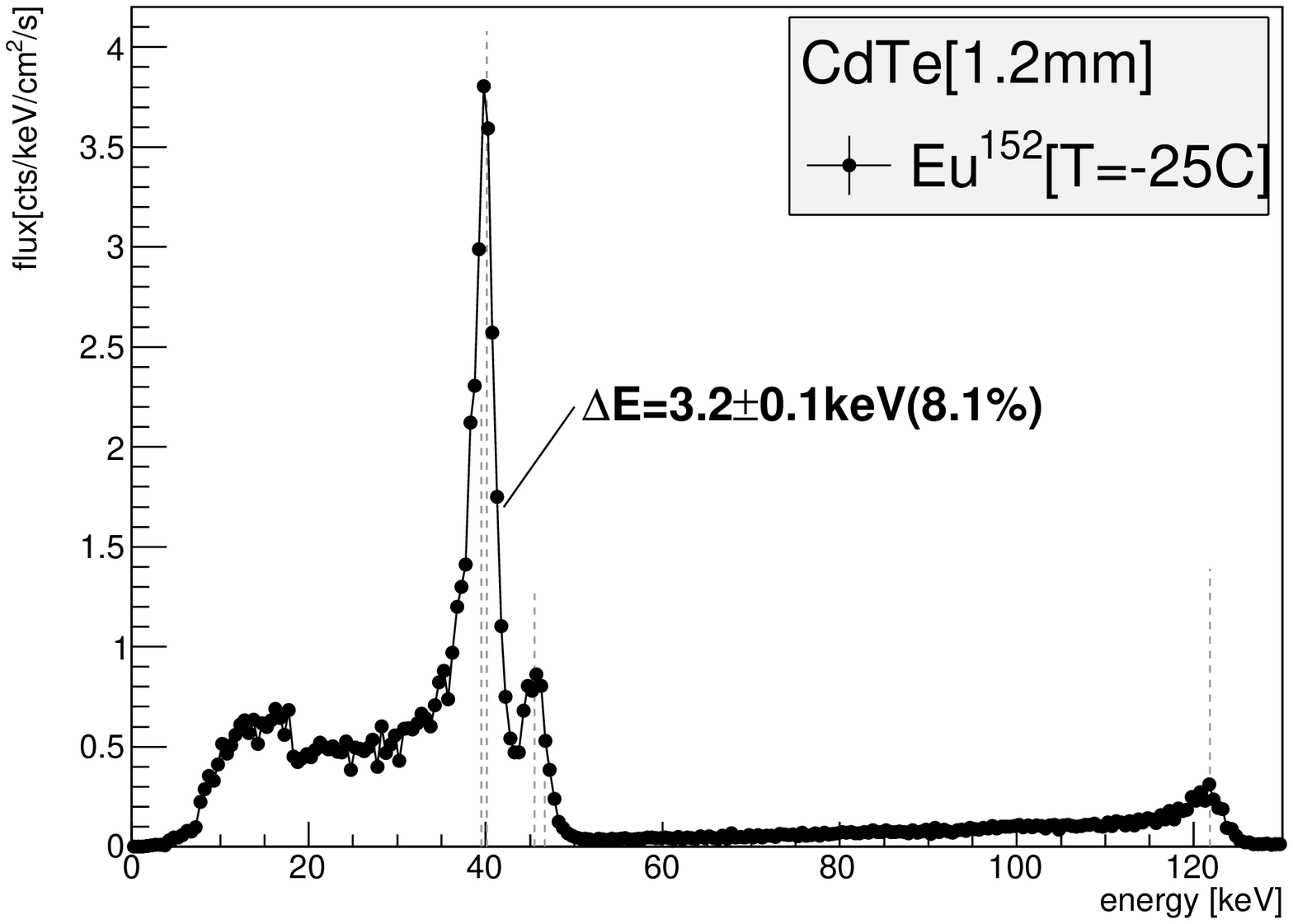}&
\includegraphics[scale=0.4]{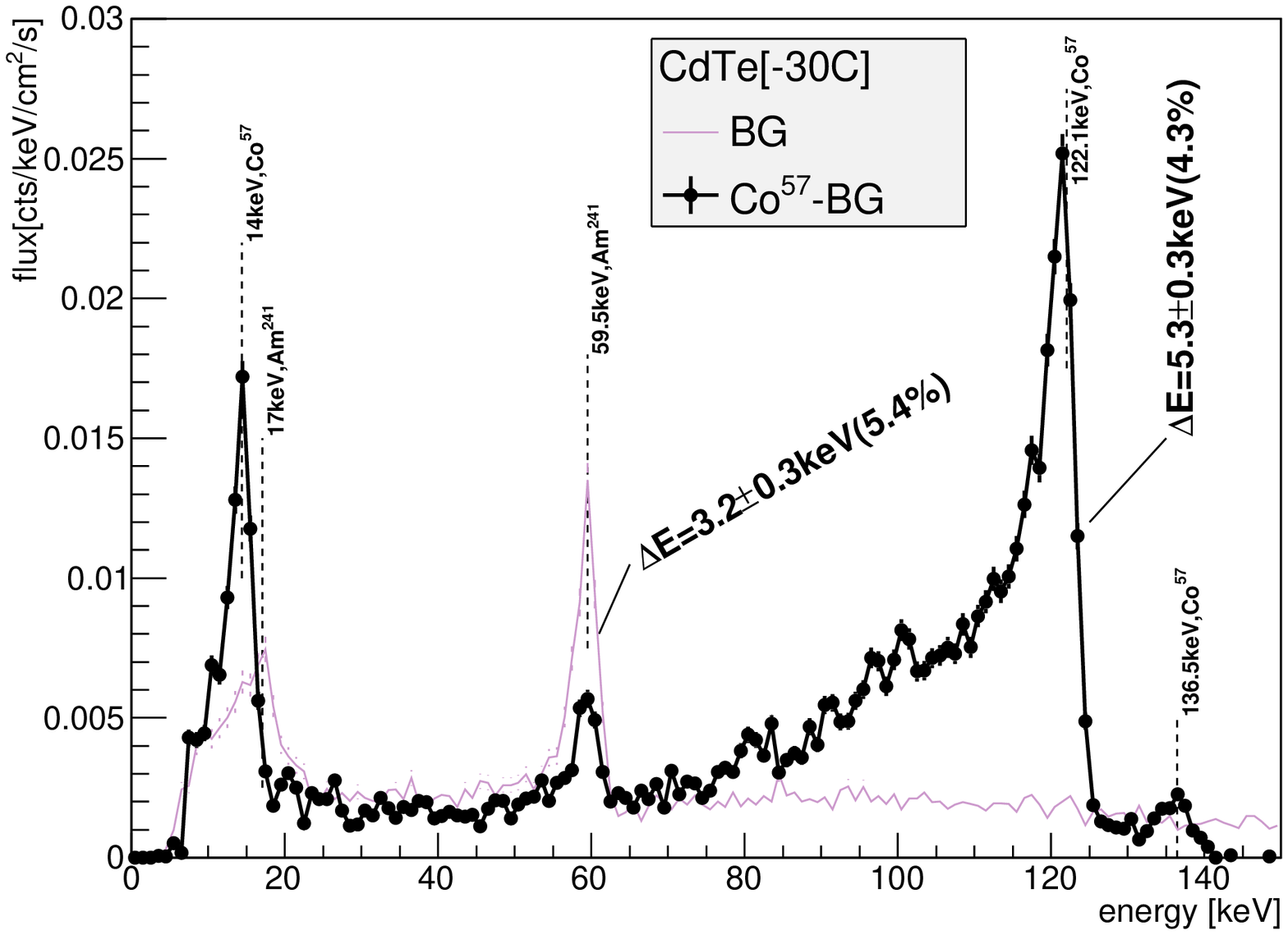}\\
\includegraphics[scale=0.7]{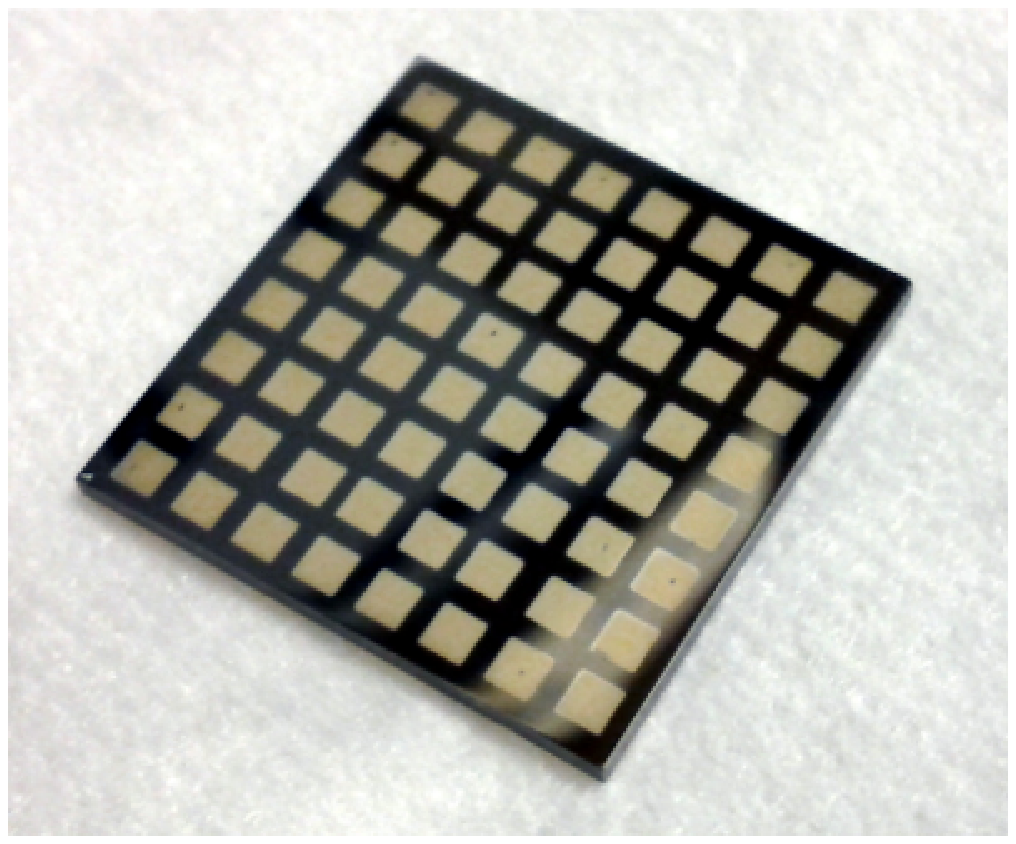}&\\
\end{tabular}

\caption{\label{fig:CdTe_SelfMade} CdTe detector ($20 \times 20 \times
1.2 \, \rm{mm}^{3}$, substrate from Acrorad). {\bf Bottom left:} Detector
fabricated at Washington University with Au cathode and In/Ti/Au pixels.
{\bf Top left:} Energy spectrum measured with $^{152}$Eu ($T=-25^{\circ} \,
\rm{C}$ and bias voltage of $-400\, \rm{V}$) with excellent energy resolution. 
The vertical lines indicate the nominal line energies (relative heights represent 
emission intensities times photo-absorption probability in $1.2 \, \rm{mm}$ of
CdTe). {\bf Top right:} Energy spectrum measured with a (weak) $^{57}$Co
source ($T=-30^{\circ} \, \rm{C}$ and bias voltage of $-400\, \rm{V}$). 
The background spectrum with lines at $17 \, \rm{keV}$ and $59.5 \, \rm{keV}$ 
is a likely result of an $^{241}$Am smoke detector installed in the temperature chamber. 
The $^{57}$Co line at $14 \, \rm{keV}$ can be resolved.}
\end{figure}

We also bought two $20 \times 20 \times 1.2 \, \rm{mm}^{3}$ CdTe
crystals from Acrorad and fabricated detectors with a Au cathode and
In/Ti/Au pixels (Fig.~\ref{fig:CdTe_SelfMade}, bottom). I/V
curves of the detectors were taken and energy spectra were recorded at
different temperatures and with different radioactive isotopes. We find
excellent performance (measured at $T=-30^{\circ} \rm{C}$) with all-pixel
averaged energy resolutions of $3.2 \pm 0.1 \, \rm{keV}$ ($8.1\%$) at
$40 \, \rm{keV}$ (Fig~\ref{fig:CdTe_SelfMade}, top left), $3.2 \pm 0.3 \,
\rm{keV}$ ($5.4 \pm 0.4\%$) at $59.5 \, \rm{keV}$ and $5.3 \pm 0.3 \,
\rm{keV}$ ($4.3 \pm 0.2\%$) at $121.8 \, \rm{keV}$. Note, that the
$^{152}$Eu $40 \, \rm{keV}$ line actually consists of two close-by lines
at $39.5 \, \rm{keV}$ and $40.1 \, \rm{keV}$. Furthermore, we resolve
the $14 \, \rm{keV}$ line of $^{57}$Co corresponding to the lowest
energy threshold of around $10 \, \rm{keV}$ achieved in our group so far
(see Fig.~\ref{fig:CdTe_SelfMade}, top right).

\section{SUMMARY}
\label{sec:summary}

In the presented paper we reported on our progress in the development 
and testing of X-ray detectors with fine-pitch pixelated 
anode. We designed a small-pixel photomask consisting of pixel blocks
with different pixel dimensions and pixel pitches. The investigated
pixel pitch range and pixel size range are: $200 - 500\, \mu$m, and 
$100 - 250\, \mu$m, respectively. The first detector, made of CZT,
with small-pixel blocks has been fabricated and is currently being
tested. The progress of the performance study of the fine-pitch pixelated
CZT detector will be reported at the SPIE 2014 conference.

Different methods of contacting pixels to the readout ASIC were considered
(pogo pins, zebra pads, direct wire bonding) and successfully tested on pixelated
anodes with $2.4\, \rm{mm}$ pixel pitch. Currently, we are in the process of 
testing how these methods can be applied to fine-pitch pixels.

We also characterised $1\, \rm{mm}$ thin CdTe detectors. We studied
their performance at different temperatures. The optimal temperature,
yielding best performance in terms of energy resolution but also 
dark current stability, was determined. The studied CdTe detectors
show excellent energy resolution of $\sim 8\%$ at $40\, \rm{keV}$.

%%%%%%%%%%%%%%%%%%%%%%%%%%%%%%%%%%%%%%%%%%%%%%%%%%%%%%%%%%%%%
\acknowledgments     %>>>> equivalent to \section*{ACKNOWLEDGMENTS}       
We are grateful for NASA funding from grant NNX13AC49G.
%%%%%%%%%%%%%%%%%%%%%%%%%%%%%%%%%%%%%%%%%%%%%%%%%%%%%%%%%%%%%
%%%%% References %%%%%

\bibliography{article_bibliography}   %>>>> bibliography data in report.bib
\bibliographystyle{spiebib}   %>>>> makes bibtex use spiebib.bst

\end{document}